\documentclass[twocolumn,pdflatex,sn-nature]{sn-jnl}

\usepackage{epstopdf}
\usepackage{epsfig}
\usepackage{graphicx}%
\usepackage{multirow}%
\usepackage{amsmath,amssymb,amsfonts}%
\usepackage{amsthm}%
\usepackage{mathrsfs}%
\usepackage[title]{appendix}%
\usepackage{xcolor}%
\usepackage{textcomp}%
\usepackage{manyfoot}%
\usepackage{booktabs}%
\usepackage{algorithm}%
\usepackage{algorithmicx}%
\usepackage{algpseudocode}%
\usepackage{listings}%
\usepackage{indentfirst}
\usepackage{hyperref}
\usepackage{subfigure}
\hypersetup{
	colorlinks=true,
	linkcolor=black,
	anchorcolor=black,
	citecolor=black,
	urlcolor=black}
\usepackage{geometry}
\newgeometry{twocolumn,top=15mm,bottom=15mm,left=15mm,right=15mm}
\usepackage{flushend}

\theoremstyle{thmstyleone}%
\theoremstyle{thmstyletwo}%

\theoremstyle{thmstylethree}%

\flushend
\begin{document}

\title[Article Title]{Movable and Reconfigurable Antennas for 6G: Unlocking Electromagnetic-Domain Design and Optimization}


\author[1]{\fnm{Lipeng} \sur{Zhu}}\email{zhulp@nus.edu.sg}

\author[2]{\fnm{Haobin} \sur{Mao}}\email{maohaobin@buaa.edu.cn}

\author[1]{\fnm{Ge} \sur{Yan}}\email{geyan@u.nus.edu}

\author[1]{\fnm{Wenyan} \sur{Ma}}\email{wenyan@u.nus.edu}

\author*[2]{\fnm{Zhenyu} \sur{Xiao}}\email{xiaozy@buaa.edu.cn}

\author*[1]{\fnm{Rui} \sur{Zhang}}\email{elezhang@nus.edu.sg}

\affil[1]{\orgdiv{Department of Electrical and Computer Engineering}, \orgname{National University of Singapore}, \orgaddress{\street{4 Engineering Drive 3}, \postcode{117583}, \country{Singapore}}}

\affil[2]{\orgdiv{School of Electronic and Information Engineering}, \orgname{Beihang University}, \orgaddress{\street{No. 37 Xueyuan Road, Haidian District}, \city{Beijing}, \postcode{100191}, \country{China}}}


\abstract{
	The growing demands of 6G mobile communication networks necessitate advanced antenna technologies. Movable antennas (MAs) and reconfigurable antennas (RAs) enable dynamic control over antenna's position, orientation, radiation, polarization, and frequency response, introducing rich electromagnetic-domain degrees of freedom for the design and performance enhancement of wireless systems. This article overviews their application scenarios, hardware architectures, and design methods. Field test and simulation results highlight their performance benefits over conventional fixed/non-reconfigurable antennas.
	
	}




\maketitle

\section{Introduction}\label{sec1}
The rapid growth and diversification of wireless services have driven the evolution of mobile communication networks from the first generation (1G) to the fifth generation (5G). As spectrum resources remain limited, enhancing spectral efficiency has always been a key research focus to meet growing data rate demands. Multiple-input multiple-output (MIMO) technology has emerged as a cornerstone in this pursuit, exploiting spatial degrees of freedom (DoFs) via multiple antennas at base stations (BSs) and/or user terminals \cite{Paulraj2004Anover}. MIMO systems provide diversity and multiplexing gains, significantly improving reliability and capacity in modern mobile communication networks. To meet increasing performance requirements, MIMO has evolved into massive MIMO in 5G \cite{larsson2014mimo} and is envisioned to scale further into extremely large-scale MIMO (XL-MIMO) in the future sixth-generation (6G) networks for more flexible beamforming and enhanced spatial multiplexing \cite{wang2024mimo, lu2024mimo}. However, these advancements also lead to increased hardware complexity, higher power consumption, and greater signal processing overhead, posing critical challenges for practical deployment and sustainability.

To address these limitations, recent efforts have explored cost-effective MIMO implementations through novel antenna architectures. In this context, movable antennas (MAs) \cite{zhu2024modeling,zhu2024magazine,ma2024mimo,shao20246DMAModel,shao20256dma} and reconfigurable antennas (RAs) \cite{zhang2022pattern,ying2024reconfigurable,heath2025tri,chen2025dbraa} have gained attraction for their ability to flexibly adjust antenna properties, including position, orientation, radiation pattern, polarization, and frequency response. MAs, also known as six-dimensional MA (6DMA) in the general form \cite{shao20246DMAModel,shao20256dma}, reconfigure external placement (position and orientation), while RAs tune internal radiation characteristics. Despite their differences, both MAs and RAs share the unique capability of reconfiguring wireless channels directly in the electromagnetic domain, complementing traditional analog/digital signal processing. This paradigm shift has sparked broad interest, revealing great promise in balancing system performance with cost and energy efficiency \cite{zhu2025tutorial,shao2025tutorial,chen2025remaa,zheng2025tri,castellanos2025embracing}.

\begin{figure*}[h]\label{fig1}
	\centering
	\includegraphics[width=16cm]{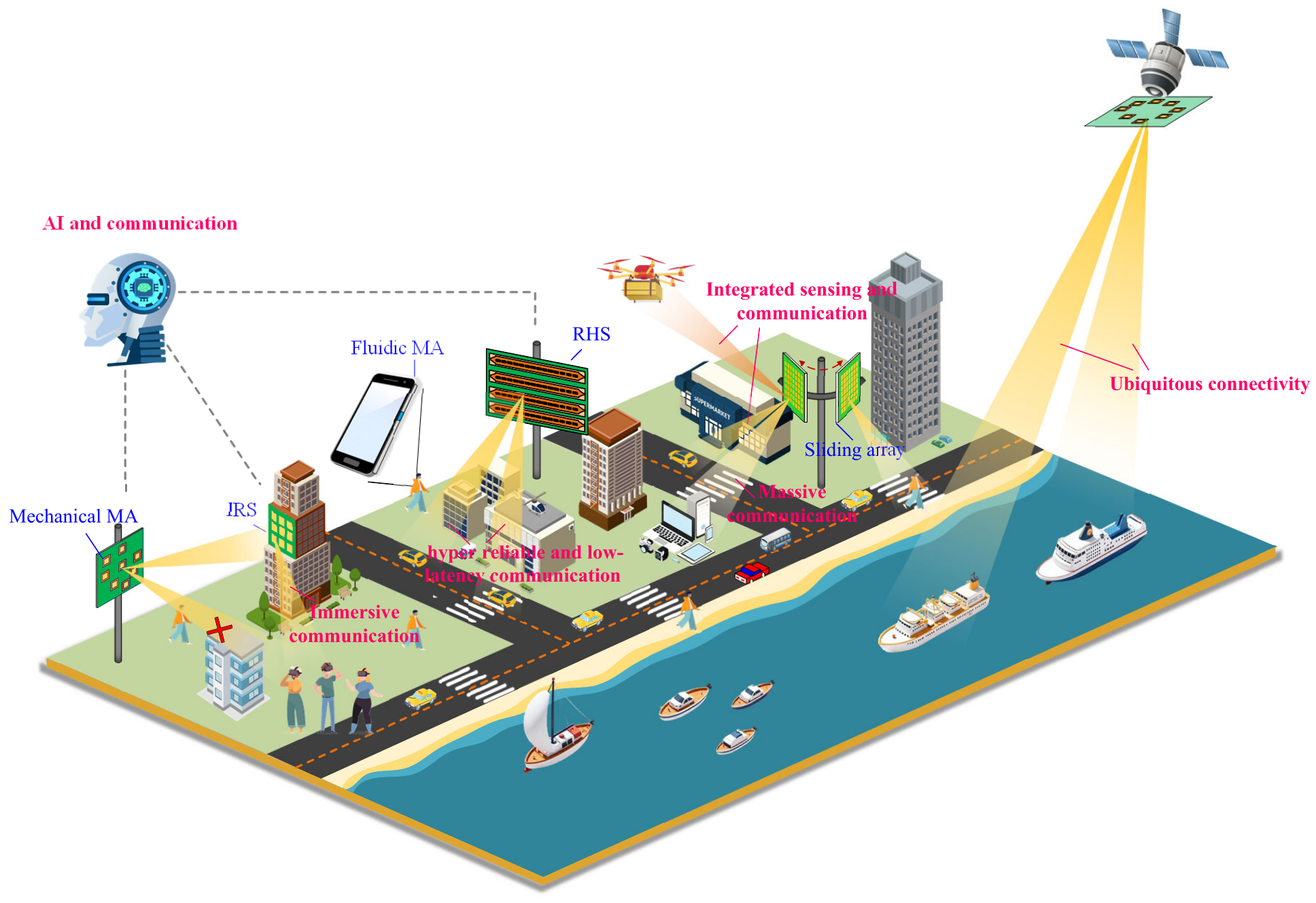}
	\caption{Typical usage scenarios for MAs and RAs towards 6G.}
\end{figure*}

As shown in Fig. \ref{fig1}, MAs and RAs enable considerable performance enhancement across different 6G usage scenarios \cite{ITU2023}. By dynamically reconfiguring wireless channels, they improve spectral efficiency and coverage for data-intensive applications such as virtual reality (VR) and Internet of Things (IoT), enhancing signal strength and mitigating interference. Their ability to adapt channel conditions also supports ultra-low latency and high reliability, which are essential for mission-critical tasks like autonomous driving. Furthermore, the flexibility in adjusting position, orientation, radiation, and polarization significantly improves sensing accuracy in dynamic environments. This reconfigurability facilitates integrated sensing and communication (ISAC), where antenna movement and configuration are jointly optimized to balance dual functions \cite{li2025isac}. Additionally, artificial intelligence (AI) can be employed to enhance real-time control, adaptation, and resource coordination in dynamic environments.

To unlock the full potential of MAs and RAs in future wireless networks, this article provides a comprehensive overview of their technical foundations, challenges, and recent progress. We first introduce key hardware architectures at both the antenna element and array levels. Then, typical design approaches for antenna movement and configuration are discussed. Field test and simulation results are presented to showcase their performance benefits. Finally, several open research problems are highlighted to guide future investigations in this emerging field.

\section{Hardware Architectures}\label{sec2}
This section categorizes the hardware implementations of MAs and RAs into element-level and array-level architectures. They can be realized through mechanical, fluidic, electronic, or hybrid means, offering varying levels of flexibility, complexity, and response speed.

\begin{figure*}[h]
	\centering
	\includegraphics[width=18cm]{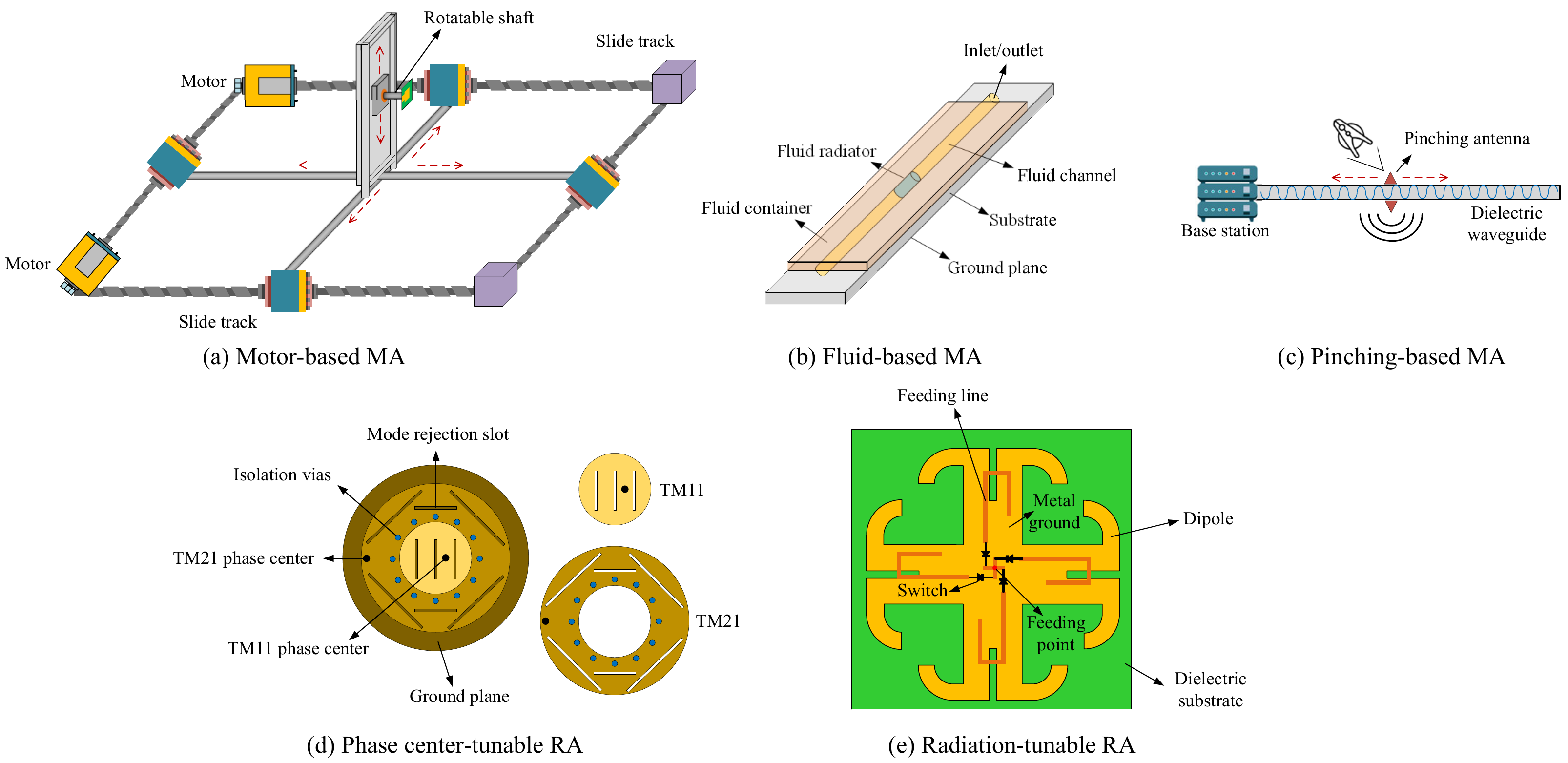}
	\caption{Hardware architectures of typical element-level MAs and RAs.}
	\label{fig2}
\end{figure*}

\subsection{Element-level Architectures}
Fig. \ref{fig2}(a) shows that the position and/or orientation of an antenna element can be controlled by motorized slides or microelectromechanical systems (MEMS) \cite{zhu2024magazine,shao20256dma}. Alternatively, fluidic actuation, as shown in Fig. \ref{fig2}(b), enables movement by adjusting the pressure in syringes or using micropumps to reposition metallic or non-metallic fluid radiators \cite{New2024fluid}. However, achieving fluidic rotation remains challenging. A recently proposed pinching antenna, illustrated in Fig. \ref{fig2}(c), uses dielectric particles (e.g., plastic pinches) to move along waveguides for large-scale repositioning \cite{ding2025pinching,liu2025pinching}.

In addition to physical movement, antenna characteristics can be reconfigured electronically. For instance, Fig. \ref{fig2}(d) depicts a dual-mode patch antenna capable of shifting its phase center by simultaneously exciting TM11 and TM21 modes \cite{ning2025movable}. Other element-level reconfiguration techniques include electronic, optical, mechanical, and smart-material-based methods \cite{ojaroudi2020reconfigurable}. A typical example is shown in Fig. \ref{fig2}(e), where switching among four positive-intrinsic-negative (PIN) diodes tunes the radiation pattern of the antenna \cite{jin2018reconfigurable}.

\subsection{Array-level Architectures}
\begin{figure*}[h]
	\centering
	\includegraphics[width=18cm]{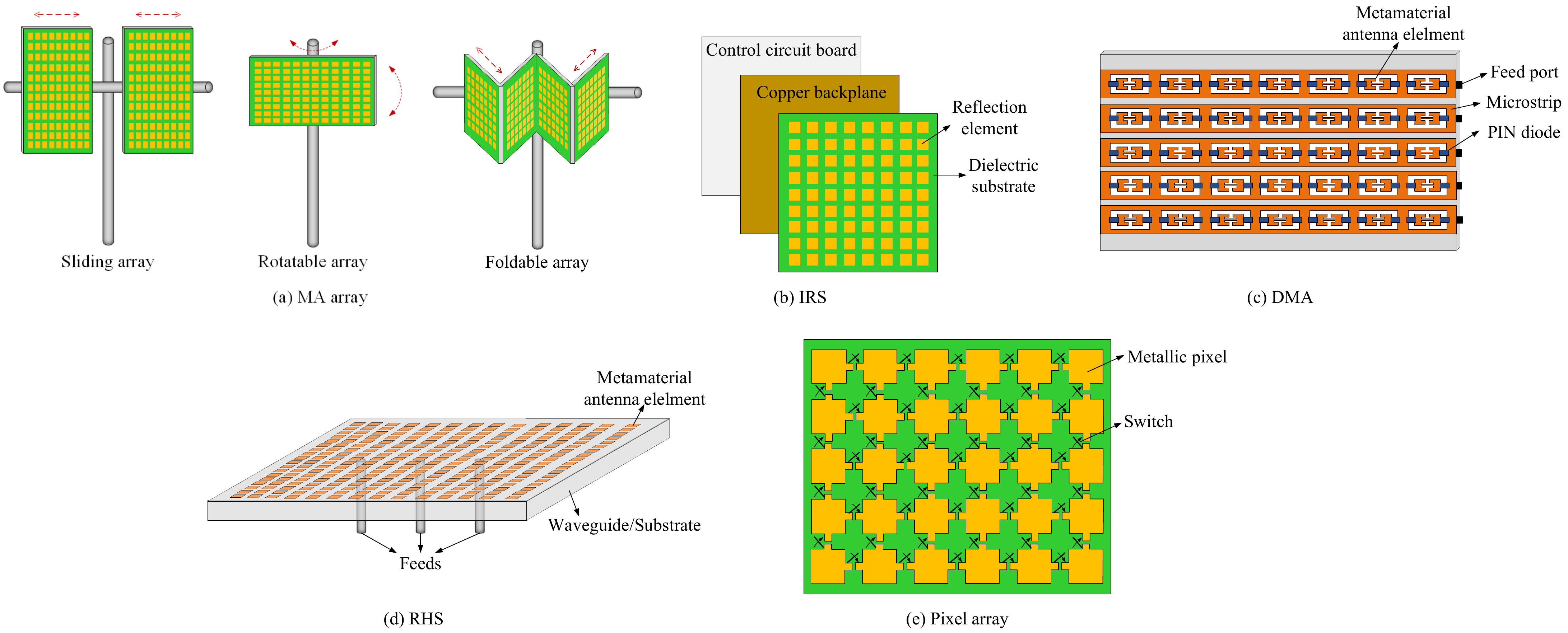}
	\caption{Hardware architectures of typical array-level MAs and RAs.}
	\label{fig3}
\end{figure*}

At the array level, MAs and RAs can jointly adjust multiple elements' positions, orientations, or electromagnetic properties. Fig. \ref{fig3}(a) presents several mechanical MA array architectures. Specifically, multiple antenna sub-arrays possess the capacity for mobility in the form of sliding, thus allowing for the enlargement of the aperture of the sliding array. This indeed promotes the creation of near-field effects and further enhances signal transmissions in both the angle and distance domains \cite{zhu2025MAnear,ning2025movable}. The rotatable array \cite{zheng2025rotatable,zheng2025rotatableMag}, which can be considered a special case of 6DMA \cite{shao20246DMAModel,shao20256dma}, allows to flexibly alter its orientation for adapting to the time-varying and non-uniformly distributed users, which helps increase the desired signal power and mitigate interference among different user clusters. The foldable array can be flexibly expanded or folded in alignment with distinct application requirements, where the former mode can be applied to enhance the array aperture and the latter can be utilized for reducing wind disturbance \cite{ning2025movable}.

Apart from these mechanically MA arrays, several emerging RA arrays have also demonstrated great potential to overcome the deficiency of conventional MIMO systems. Fig. \ref{fig3}(b) shows a passive intelligent reflecting surface (IRS) with three layers: reconfigurable metallic patches, a copper backplane to prevent signal leakage, and a control board for dynamic tuning \cite{wu2021irs}. A dynamic metasurface antenna (DMA), shown in Fig. \ref{fig3}(c), features microstrips with programmable metamaterial elements to form desired beams \cite{shlezinger2021dma}. A reconfigurable holographic surface (RHS), illustrated in Fig. \ref{fig3}(d), uses embedded feeds and waveguides to radiate amplitude-controlled reference waves via tunable elements \cite{deng2021rhs,gong2024hmimo}. Fig. \ref{fig3}(e) shows a pixel array, which alters antenna geometry by switching connections between adjacent metallic pixels \cite{lotfi2017pixel}.

\subsection{Comparison and Hybrid Designs}
It is imperative to deliberate on the distinctive advantages and limitations of the above hardware architectures of MAs and RAs for practical implementation in future 6G mobile communication systems. For instance, mechanical MAs can be moved over a wide range and are of fine movement resolution \cite{zhu2025tutorial,fu2025extremely}, while their movement is generally constrained by kinematic factors, such as acceleration and speed, which may prevent them from quick response to rapid variation of instantaneous channels. In comparison, electronic RAs possess the capacity to respond expeditiously to highly dynamic channel conditions, and they can be fabricated to be highly compact in size. However, the state and resolution of electronic reconfiguration are in general limited due to cost considerations. 

Additionally, array-level MAs and RAs enable the collective adjustment of multiple antennas' locations, orientations, and electromagnetic properties. This helps achieve a trade-off between performance degradation and reduced hardware costs \cite{zhang2025hybridMA,zhu2025multiuser}. Nevertheless, hybrid architectures can combine the advantages of different hardware architectures while mitigating their respective drawbacks. For example, electronic RA elements can be formed within a mechanical MA array to provide superior flexibility for reconfiguring channel conditions. Specifically, large-scale location adjustments can be achieved through movement of the mechanical array to adapt to slow-varying statistical channel conditions, while small-scale position/radiation alterations can be achieved through tuning of the electronic elements to adapt to fast-varying instantaneous channel conditions. Similarly, other aforementioned hardware architectures of MAs and RAs can also be tailored for more effective hybrid designs for practical implementations and specific applications.

\subsection{Deployment Considerations}
In light of the divergent deployment requirements across different scenarios, it is important to select appropriate hardware architectures to ensure a balanced system performance and implementation overhead. For example, BSs are adequately supplied with energy and have sufficient space for antenna array installation, which facilitates the implementation of mechanical array-level MAs for performing position/orientation adjustments to accommodate long-term statistical channels, switch between communication and sensing services, or support aerial and terrestrial connectivity. Meanwhile, to further enhance system performance, electronic element-level RAs can be employed to achieve real-time reconfiguration of instantaneous channel conditions. Additionally, it is more likely to mount MEMS-enabled MAs, fluid-based MAs, or electronic RAs on user terminals to guarantee the compactness and portability. Furthermore, it is appealing to deploy passive counterparts in environments to reconfigure the signal propagation conditions. For example, passive IRS can be mounted on the building facades to help enhance the desired signal strength and suppress undesired interference via joint movement and radiation optimization. 

It is foreseeable that future wireless networks will operate over different frequency bands for diverse applications, so different requirements will arise for the selection of antenna hardware architectures. For example, the wireless channels over sub-6 GHz bands undergo rich scattering, which require real-time antenna movement or configuration to adapt to small-scale fading channels. In this context, MA/RA elements shall be applied to exploit the large numbers of multi-path components and achieve significant spatial diversity gains. In comparison, MA/RA arrays may be more compatible with the wireless channels over millimeter-wave (mmWave)/terahertz (THz) bands that experience path sparsity, where more flexible beamforming and enhanced spatial multiplexing can be achieved, benefiting from the geometry/radiation flexibility.

\section{Antenna Movement/Configuration Design}\label{sec3}
The antenna movement and configuration design of MA- and RA-enabled systems are pivotal but challenging to harness their full potential and benefits. Based on the reliance of channel state information (CSI), existing approaches can be classified into three main categories, i.e., CSI-based, CSI-free, and AI-driven methods, which are outlined as follows.

\subsection{CSI-based Optimization}
CSI-based optimization involves a preliminary step of acquiring the complete knowledge of CSI \cite{ma2023MAestimation,xiao2023channel}, which is followed by the antenna movement and configuration design. The best way is to derive a closed-form solution for antenna position, orientation, or electromagnetic properties under specific system setups, e.g., those in \cite{zhu2023fullgain,ma2024sensing,shao2025channel}. 
However, it is noted that obtaining an optimal closed-form solution for antenna movement or configuration may be challenging in general due to the highly non-linear metric function and highly coupled variables in practical systems. In this regard, performing an exhaustive search over discrete solutions and selecting the ones achieving the best system performance would be another straightforward method \cite{shao20256dmadiscrete}. However, this method may incur high computational complexity and energy consumption, as candidate solutions must be densely sampled to guarantee optimality. This significantly impedes its practical implementation in real-world scenarios. Although evolutionary algorithms with heuristic searching capabilities can be similarly leveraged to pursue favorable antenna positions, orientations, or configurations, such as particle swarm optimization \cite{xiao2024multiuser}, firefly algorithm \cite{hoang2024firefly}, and hippopotamus optimization \cite{xiao2024noma}, it may still exhibit low efficiency when confronted with substantial numbers of antenna elements or ultra-large movable regions. To overcome this limitation, gradient descent/ascent \cite{zhu2024multiuser,hu2024secure}, successive convex approximation (SCA) \cite{ma2024mimo,feng2024sumrate}, semi-definite relaxation (SDR) \cite{wu2019irs}, and other non-convex programming based techniques have been adopted to obtain sub-optimal antenna setups, while they are prone to being trapped in local optima. Moreover, to further decrease the complexity of optimizing MAs' positions within a continuous moving region, several discrete sampling based schemes have been developed to achieve optimal on-grid position selection, such as graph-based approach \cite{mei2024graph} and orthogonal matching
pursuit (OMP)-based method \cite{yang2024flexible,yang2025wmmse}. Nevertheless, these methods fall short of achieving truly best system performance since the DoFs in the continuous spatial domain have not been fully exploited. Therefore, the off-grid optimization should be adopted to refine the on-grid positions of the antennas into continuous ones, thereby improving the overall system performance.

Notably, the aforementioned optimization methods can be implemented based on either instantaneous or statistical CSI. On one hand, antenna movement/reconfiguration based on instantaneous CSI can yield the best performance but necessitates frequent adjustments in fast fading channels, potentially incurring substantial pilot overhead and energy consumption. On the other hand, the statistical CSI remains unchanged over a long time period, and the antennas require repositioning or reconfiguration only upon changes of the statistical CSI or user distributions, thereby significantly reducing system overhead at the cost of certain performance loss \cite{chen2025MAstatistical,yan2025movable}.

\subsection{CSI-free Design}
In contrast to CSI-based optimization, CSI-free design does not require complete knowledge of CSI to design the antenna position, orientation, or configuration. This avoids the estimation of numerous channel coefficients and thus mitigates both the pilot overhead and the CSI estimation complexity. Typical CSI-free optimization approaches include codebook-based methods and adaptive optimization based on online measurement. Specifically, the codebook-based approach first involves designing a codebook with several antenna modes offline, guaranteeing communication performance under different channel conditions. Subsequently, online training between the BS and users is performed, wherein the BS evaluates the system performance under different antenna modes. Finally, the mode that achieves the best system performance is selected for wireless transmission. Despite investigations in \cite{zheng2025tri,an2024codebook,hwang2024codebook}, the codebook-based design for MAs and RAs still remains a significant challenge since the antenna movement and configuration are generally coupled with analog/digital signal processing parameters. In contrast, the adaptive optimization based on online measurement treats the design for antenna movement/configuration as a black-box optimization problem, which can be then efficiently addressed by methods such as Bayesian optimization \cite{shahriari2016bayesian} and zeroth-order optimization \cite{liu2020zeroth} with previous measurements. For example, the authors in \cite{zeng2025derivative} maximized the sum-rate of multiple users in an MAs-enabled communication system via a derivative-free approach. In their proposed method, the gradient of the objective function was approximated based on the previously received signals, which alleviates the need for complete CSI acquisition.

\subsection{AI-driven Method}
Beyond the CSI-based optimization and CSI-free design, AI-driven method paves a new way for MA and RA system design. On one hand, the supervised machine learning method is capable of extracting the relationship between the CSI and the optimal antenna setups by offline training \cite{zhu2024magazine}. Then, the real-time channel measurements can be input to the well-trained neural networks, and the favorable antenna setups will be output to realize the system's overarching performance goals. However, this may be challenging in real-world applications due to the absence of labeled training data. In addition, the neural networks should be retrained when certain prior system parameters change, such as the number of antenna elements and the size of the moving region, which challenges the algorithm's adaptivity and efficiency. In light of these considerations, unsupervised machine learning methods, such as reinforcement learning and deep reinforcement learning, can be leveraged to optimize the antenna movement and configuration, bypassing the need for labeled training data \cite{wang2024ai}, which generally comprises an offline training phase and an online application phase. Specifically, during the offline training phase, the agents interact with the environment to learn policies for antenna movement and configuration by repeatedly taking actions and updating states. Consequently, the primary objective function increases in line with the reward function during the training process. Based on the obtained optimal policies during the offline training process, online solutions for antenna movement and configuration can then be efficiently acquired. Despite the flexibility and effectiveness of the approaches in \cite{weng2024learning,yang2021learning,adhikary2023holographic}, significant pilot signaling overheads may be incurred prior to the training process in the aforementioned explicit CSI-based AI methods. On the other hand, CSI-free reinforcement learning frameworks could be effective for refining the movement and configuration of MAs and RAs at relatively low cost. Looking forward, optimal system setups may be output by observing or measuring the objectives at several candidate discrete solutions, which requires careful design of the measurement setup and learning mechanism. In addition, other advanced AI-based techniques such as federated learning \cite{ahmadzadeh2025federated}, generative adversarial networks \cite{creswell2018gan}, and large language models \cite{wang2025llm} are also expected to achieve more efficient and intelligent designs for MA- and RA-aided wireless networks.

\section{Experimental Results}\label{sec4}
To validate the performance advantages of MAs and RAs, this section presents the field-test and simulation results in single-input and single-output (SISO) and multiuser multiple-input and single-output (MISO) systems, respectively.
\subsection{Field Test for SISO System}
\begin{figure*}[h]
	\centering
	\subfigure[Prototype]{\includegraphics[width=8.5cm]{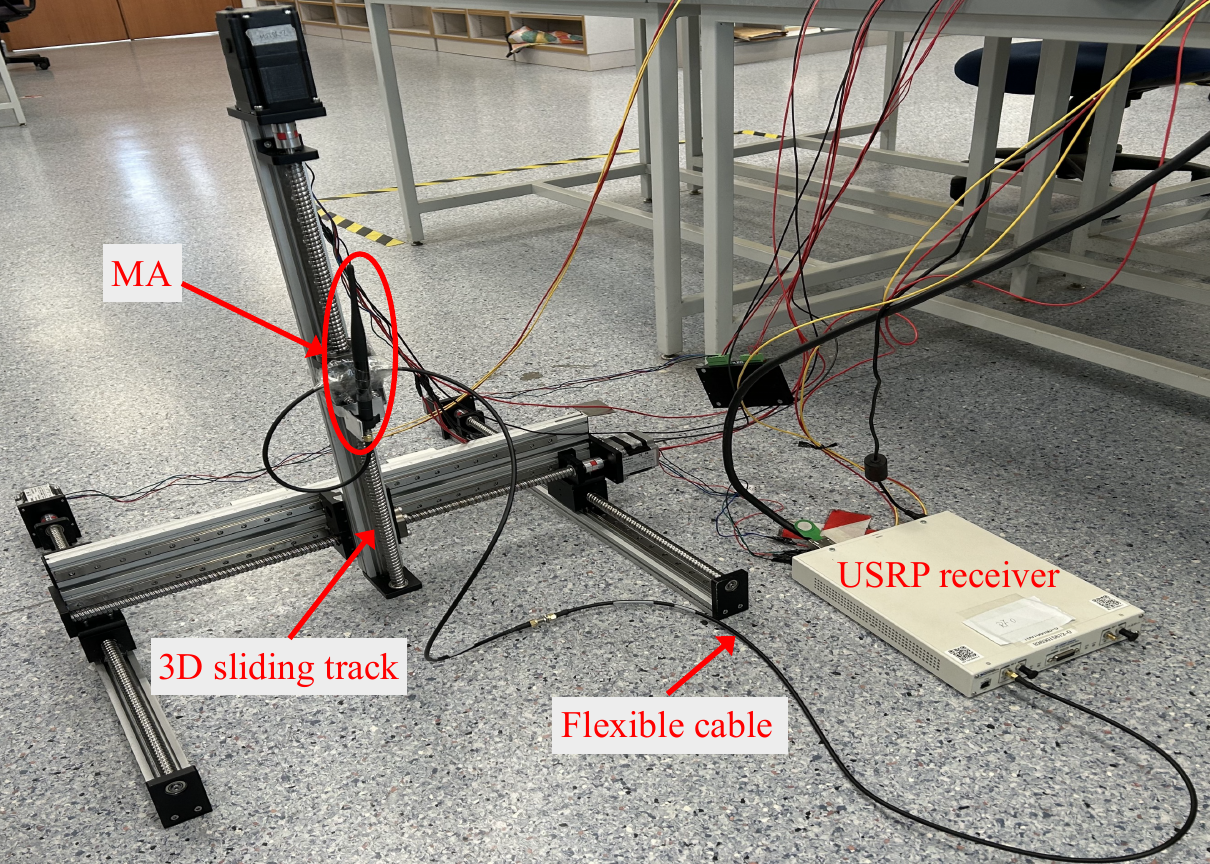} \label{Fig_prototype}}
	\subfigure[Received SINR versus the size of the antenna moving region]{\includegraphics[width=8.5cm]{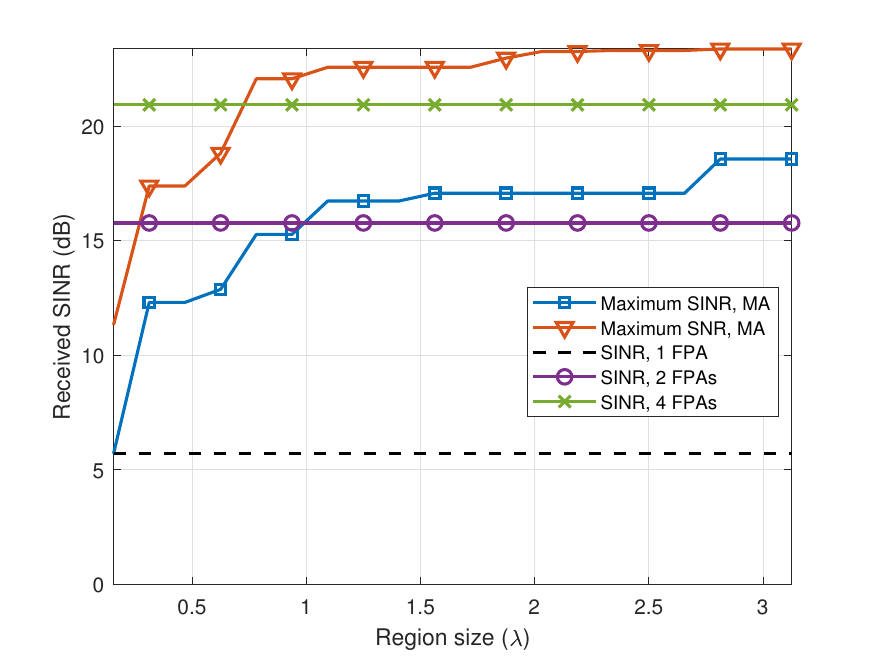} \label{Fig_SISO}}
	\caption{Prototype of the MA-aided SISO receiver and field-test results. (Experimental setup: The system consists of an FPA-based transmitter, an FPA-based jammer, and an MA-based receiver. The carrier frequency is 2.49 GHz (i.e., wavelength $\lambda=12$ cm). The maximum antenna moving region is a 3D cube of size $3.125\lambda=37.53$ cm, which is divided into equally spaced $25 \times 25 \times 25$ grids. The optimal MA position is selected within a 3D subspace (with varying size) of the moving region for maximizing the received SNR or SINR. For multi-FPA receivers, the antennas are spaced by distance $\lambda/2$, and the optimal minimum mean square error (MMSE) beamforming is employed for maximizing the received SINR.)}
	\label{SISO}
\end{figure*}
To validate the performance gain in improving received signal power and in reducing undesired interference, we develop an MA-aided SISO prototype shown in Fig. \ref{SISO} (a). With the aid of sliding tracks, the MA position can be flexibly adjusted in the three-dimensional (3D) space. A performance comparison between the MA and fixed-position antenna (FPA) systems is presented in Fig. \ref{SISO} (b). As can be observed, by optimizing the antenna position at the receiver, the achieved signal-to-interference-plus-noise ratio (SINR) for the MA receiver improves over 10 dB as the size of the antenna moving region increases, which even outperforms multi-FPA systems adopting digital beamforming. Moreover, for larger antenna moving regions, the gap between the received SINR and signal-to-noise ratio (SNR) decreases, indicating that the MA can efficiently leverage the channel spatial variation for interference mitigation.

\subsection{Simulation for Multiuser MISO System}
\begin{figure*}[h]
	\centering
	\subfigure[Simulation scenario]{\includegraphics[width=8.5cm]{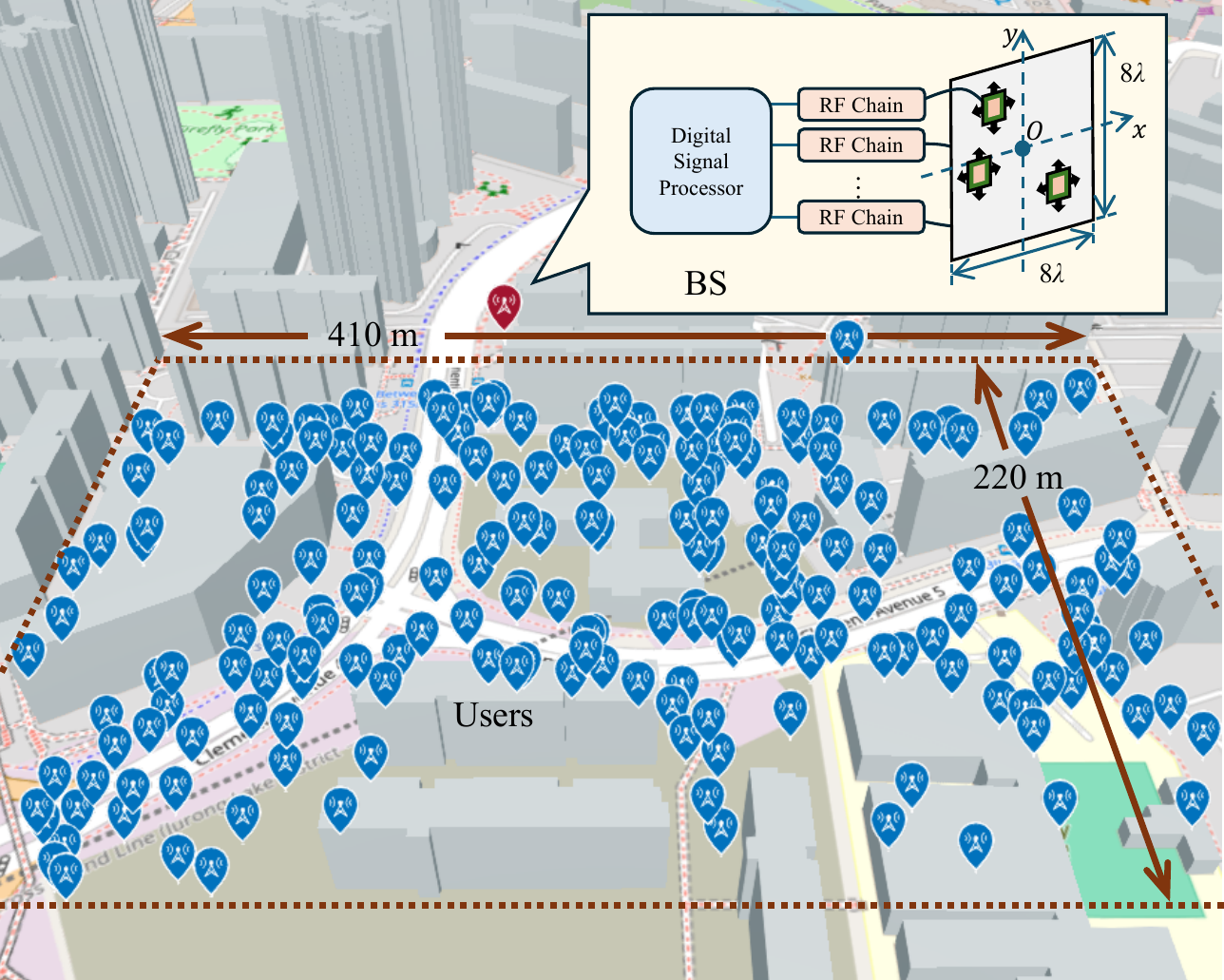} \label{Fig_raytracing}}
	\subfigure[Ergodic sum rate versus the number of users]{\includegraphics[width=8.5cm]{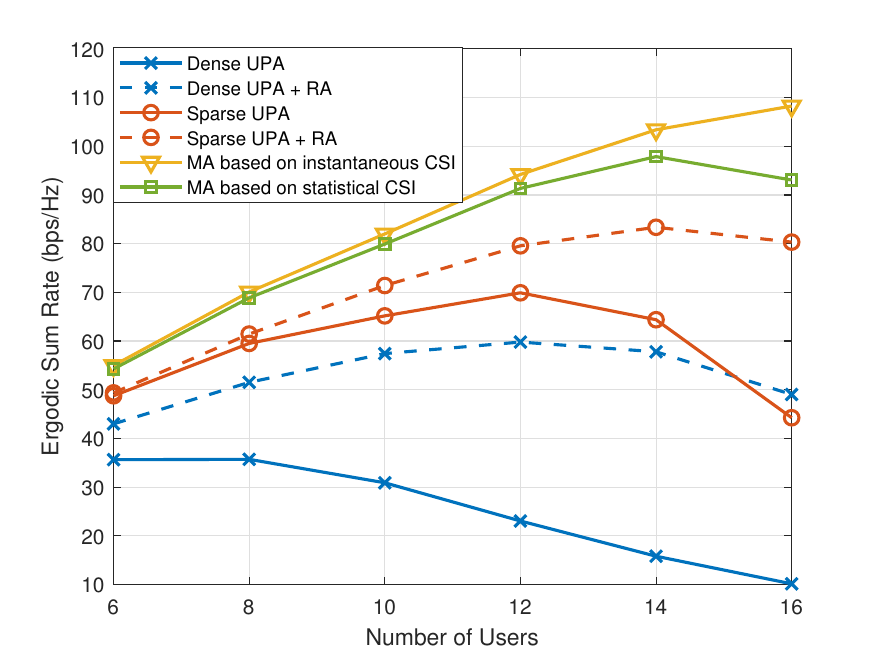} \label{Fig_MISO}}
	\caption{Multiuser MISO communication scenario and simulation results based on ray-tracing. (Simulation setup: The single-antenna users are randomly distributed within an urban area of size $220 \text{m} \times 410 \text{m}$ at Clementi, Singapore. Each channel realization is obtained via ray-tracing based on the randomly generated user locations within the target area. The dense uniform planar array (UPA) and sparse UPA schemes adopt $4 \times 4$ FPAs with inter-antenna spacing given by $\lambda/2$ and $2\lambda$, respectively, where $\lambda$ denotes the carrier wavelength and the carrier frequency is $5$ GHz. For MA systems, the BS is equipped with 16 MAs with the size of the two-dimensional (2D) antenna moving region given by $8\lambda \times 8 \lambda$, where the antenna positions are optimized based on gradient ascent for maximizing the ergodic sum rate of users \cite{yan2025movable}. Each RA has 4 candidate radiation patterns \cite{zheng2025tri}, with the optimal one selected via alternate refinement for each antenna through exhaustive search. For all schemes, zero-forcing (ZF) beamforming is adopted with the optimal power allocation determined by water-filling.)}
	\label{MISO}
\end{figure*}
Fig. \ref{MISO} shows the MA- and RA-aided multiuser MISO communication system and evaluate their performance gain compared to conventional FPAs. It is observed that both the MA and RA schemes can significantly improve the achievable average/ergodic sum-rate of users. In particular, the MA scheme with instantaneous channel-based antenna position optimization yields the best performance, which demonstrates the importance of array geometry reconfiguration in reducing multiuser channel correlation for interference mitigation. However, due to the slow movement speed of MAs, the instantaneous channel-based design may not be efficient for high-mobility users undergoing fast-fading channels. Therefore, the statistical channel-based MA position optimization turns to be a viable solution, which significantly reduces the antenna movement overhead at a certain sacrifice of the rate performance. In comparison, the RA scheme can achieve a moderate performance improvement compared to its FPA counterpart under both dense and spare UPA setups. Despite the fast response speed, the performance of RA systems is limited by the finite state of radiation patterns. It can be expected that as the number of candidate reconfigurable patterns increases, the RA system can achieve better communication performance at the cost of higher hardware complexity.

\section{Open Problems}
Despite the significant performance advantages of MAs and RAs for next-generation wireless communication systems, several open problems remain to fully unlock their potential in realizing versatile and intelligent 6G networks, as elaborated in the following.

\subsection{Advanced Antenna Architecture}
Different antenna architectures possess distinct characteristics and impose varying practical constraints on flexibility, response speed, hardware cost, and energy consumption. Concurrently, diverse wireless applications and deployment environments impose unique requirements on antenna structures. For instance, the promising ISAC framework envisioned for 6G necessitates that BSs dynamically switch between sensing and communication functions, thus demanding antenna architectures capable of adaptive function reconfiguration. Consequently, the development of advanced antenna architectures is critical to achieve an optimal balance between hardware cost and system performance. 

\subsection{Unified Modeling and Analysis}
The majority of existing investigations on MAs and RAs enabled wireless systems have been undertaken independently, where the performance advantages of their integrated design are unexplored. In this context, a unified channel model capturing the effects of both antenna movement and electromagnetic configuration remains imperative, which can then act as the foundation to facilitate the joint optimization. Furthermore, characterizing network performance limits based on the unified model becomes increasingly urgent to unleash their envisioned potential in improving wireless communication and/or sensing performance, while the extant literature on this topic is still limited.

\subsection{Efficient Algorithm Development}
Since the antenna movement and configuration variables are usually highly coupled with other system parameters, it is generally challenging to obtain globally optimal solutions. Despite promising approaches for antenna movement/configuration outlined before, there is an urgent need to develop more efficient algorithms for jointly optimizing the antenna movement and configuration to fully harness their benefits in 6G mobile communication systems. Furthermore, computationally efficient algorithms that guarantee solution quality, such as distributed optimization methods, warrant significant future research efforts to enable the practical deployment of MAs and RAs. In addition, robust optimization methods under imperfect CSI are also paramount to enhance the resilience and reliability of wireless communication and sensing systems.

\subsection{Prototyping and Experiments}
Although several prototypes have been developed to validate the performance advantages of MAs and RAs in wireless systems, the majority of them remain single-antenna implementations, with movement and rotation capabilities through mechanical, electrical, or fluidic control \cite{wang2024movable,dai2025demo,shen2024design}. To facilitate a more profound comprehension of MAs and RAs, further research is required to explore multiple MAs and RAs prototyping with hybrid control mechanisms. Concurrently, experiments across diverse application scenarios remain essential to validate the performance benefits of MAs and RAs in spatial multiplexing and flexible beamforming, providing effective insights into practical network design and deployment in the 6G era.

\subsection{Standardization and Commercialization}
Given that MA and RA technologies remain nascent, their standardization and commercialization efforts are still in their infancy, with critical areas such as unified modeling, movement/configuration management, channel estimation, and hardware design largely unexplored. However, establishing general and standardized frameworks is essential to transition MA and RA technologies from theoretical research to practical deployment. These frameworks form the foundation for widespread commercialization, enabling significant economic benefits across the wireless ecosystem. Therefore, collective efforts from academic researchers, industrial practitioners, and other stakeholders are deserved to advance this burgeoning field.

\section{Conclusion}\label{sec13}
MAs and RAs offer a transformative approach to overcoming the limitations of conventional MIMO systems by introducing reconfigurability directly in the electromagnetic domain. Through dynamic adjustment of antenna position, orientation, radiation pattern, polarization, and frequency response, MAs and RAs provide new DoFs for enhancing communication and sensing performance in 6G wireless networks. This article has reviewed their representative application scenarios, hardware architectures, and key design methodologies, along with field test and simulation validations. While promising performance gains have been demonstrated, several open challenges remain, particularly in hardware design, unified modeling, control algorithms, and system integration. Continued interdisciplinary research, spanning electromagnetics, antenna theory, communication theory, signal processing, and AI, is essential to fully unlock the potential of MA- and RA-enabled wireless systems and realize their envisioned role in next-generation intelligent and adaptive wireless networks.

%
%
%
%
%
%
%

\section*{Declarations}


\begin{itemize}
\item \emph{Funding Statement}: This work was supported by the National Natural Science Foundation of China (NSFC) under grant number 61827901 and the National University of Singapore under Research Grant A-8003646-00-00.
\item \emph{Competing Interests}: The authors declare no competing interests.
\item \emph{Author Contribution}: L. Zhu, Z. Xiao, and R. Zhang conceived the outline; L. Zhu and H. Mao wrote the paper; G. Yan and W. Ma conducted field tests and simulations.
\end{itemize}

\bibliography{movable-and-reconfigurable-antennas}

\end{document}